\documentclass[showpacs,twocolumn,superscriptaddress,amsmath,amssymb,preprintnumbers,floatfix]{revtex4}

\usepackage{amsmath,amssymb}
\usepackage{bm}
\usepackage{graphicx}
\usepackage{ae}
\usepackage{epsfig}
\usepackage{dcolumn}

\newcommand{\kpe}{\mathbf{k}\!\cdot\!\mathbf{p}\,}

\begin{document}

\preprint{(accepted at Appl.\ Phys.\ Lett.)}

\title{Band gap and band parameters of InN and GaN from quasiparticle energy
       calculations based on exact-exchange density-functional theory}

\author{P. Rinke}
\affiliation{%
 Fritz-Haber-Institut der Max-Planck-Gesellschaft, Faradayweg 4--6,
 D-14195 Berlin, Germany
}
\author{A. Qteish}
\affiliation{%
Department of Physics, Yarmouk University, 21163-Irbid, Jordan
}
\author{M. Winkelnkemper}
\affiliation{%
 Fritz-Haber-Institut der Max-Planck-Gesellschaft, Faradayweg 4--6,
 D-14195 Berlin, Germany
}
\affiliation{%
Institut f\"ur Festk\"orperphysik, Technische Universit\"at Berlin,
Hardenbergstra{\ss}e 36, D-10623 Berlin, Germany
}

\author{D. Bimberg}
\affiliation{%
Institut f\"ur Festk\"orperphysik, Technische Universit\"at Berlin,
Hardenbergstra{\ss}e 36, D-10623 Berlin, Germany
}
\author{J. Neugebauer}
\affiliation{%
Max-Planck-Institut f\"ur Eisenforschung, Department of
               Computational Materials Design, D-40237 D\"usseldorf, Germany
}
\author{M. Scheffler}
\affiliation{%
 Fritz-Haber-Institut der Max-Planck-Gesellschaft, Faradayweg 4--6,
 D-14195 Berlin, Germany
}

\date{\today}

\begin{abstract}
We have studied the electronic structure of InN and GaN employing $G_0W_0$ 
calculations based on exact-exchange density-functional theory. 
For InN our approach predicts a gap of 0.7 eV. Taking the Burnstein-Moss 
effect into account, the increase of the apparent quasiparticle gap with 
increasing electron concentration is in good agreement with the observed blue 
shift of the experimental optical absorption edge. Moreover, the concentration 
dependence of the effective mass, which results from the non-parabolicity of 
the conduction band, agrees well with recent experimental findings. Based on 
the quasiparticle band structure the parameter set for a $4\times4$ $\kpe$ 
Hamiltonian has been derived. 
\end{abstract}


%


\maketitle

The group III-nitrides AlN, GaN and InN and their alloys
have become an important class of semiconductor materials, in
particular for use in optoelectronic devices such as green and blue
light emitting diodes (LEDs) and lasers. 
Among the three materials InN is still the least
explored, due to difficulties in synthesizing high quality single crystals. 
Only very recently these 
problems have been overcome \cite{Walukiewicz/review:2006}, but 
many of the key band parameters have not been conclusively determined until now
\cite{Walukiewicz/review:2006,Vurgaftman/Meyer:2003}. 
The most controversially discussed parameter is currently still the fundamental 
band gap of InN. 
For many years it was believed to be approximately 1.9 eV, but
essentially any value between 0.65 and 2.3 eV has been 
reported in the literature over the last 30 years
\cite{Walukiewicz/review:2006}. 
However, more recent experiments on high quality samples grown by molecular 
beam epitaxy (MBE) and recent {\it ab initio} calculations 
support a significantly lower value around 0.7 eV
\cite{Davydov/etal:2002,Wu/etal:2002_1,Nanishi/etal:2003,
      Sher/etal:1999,Bechstedt/Furthmueller:2002}.

Different hypotheses have been proposed to explain the large
variation in the measured band gaps. 
Defects could be responsible for inducing states in the band gap or give rise to a
pronounced Burnstein-Moss effect due to a shift in the Fermi level caused 
by a high intrinsic electron density.
Non stoichiometry may increase the defect concentration or alter the crystaline
structure. The formation of oxides and 
oxynitrides would increase the band gap, whereas the precipitation of 
In clusters leads to additional features in optical absorption 
spectra  \cite{Walukiewicz/review:2006,Butcher/Tansley:2005}. 
In this article we demonstrate that first principles calculations can contribute 
to the solution of this fundamental question. 
By combining density-functional theory (DFT) with
many-body perturbation theory in the $G_0W_0$ approximation \cite{Hedin65}, 
which is currently the method of choice for calculating
quasiparticle excitations in solids 
\cite{Onida/Reining/Rubio:2002,Rinke/etal:2005},  
we combine atomistic control over the material with accurate calculations for
the band structure and the band gap of stoichiometric and 
defect free structures. 

\begin{figure}
  \begin{center} 
    \epsfig{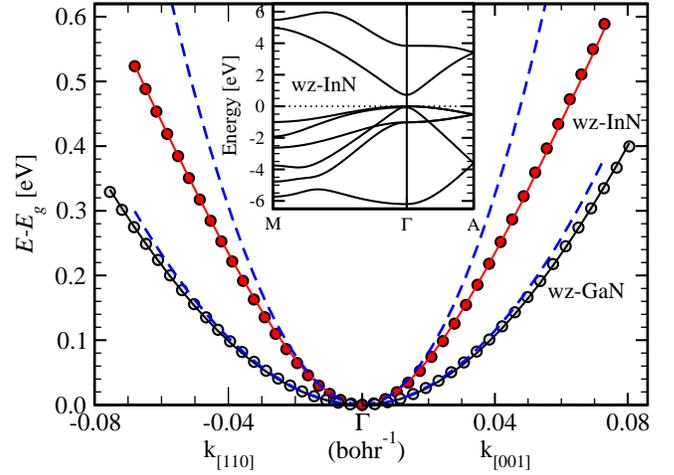}
    \caption{\label{fig:InN-CB} 
             Conduction band of wurtzite InN and GaN aligned at the bottom of
	     the conduction band: the circles are the $G_0W_0$ results,
	     the solid lines the $\kpe$ fit using Eq. \ref{eq:CB_kp}, 
	     and the dashed lines
	     the effective mass band. The inset shows the band structure of wurtzite InN.
	     }
  \end{center}
\end{figure}

Previous {\it ab initio} studies were aggravated by the fact that DFT 
calculations in the local-density approximation (LDA) predict InN to be 
metallic in the zinc blende and wurtzite structures. Subsequent $G_0W_0$ calculations only open the gap to
0.02 - 0.05 eV \cite{Usuda/Hamada/etal:2004,Kotani/Schilfgaarde:2002},
while adding self-interaction corrections to the DFT calculations, either in the
screened-exchange \cite{Sher/etal:1999}, 
exact-exchange optimized effective potential approach (OEPx) \cite{Qteish/etal:2005}, or 
self-interaction corrected (SIC) LDA approach \cite{Stampfl/etal:2000,Vogel/etal:1997},
yield a semiconductor with a band gap  
of 0.8, 1.0 and 1.6 eV for the wurtzite phase, respectively.
Here we apply the $G_0W_0$ corrections to OEPx ground state calculations, 
which are fully self-interaction free.      
We have previously shown that this approach
yields band gaps for II-VI compounds and GaN in very good agreement with
experiment \cite{Rinke/etal:2005} and for wurtzite InN our value of 0.7 eV 
(see Tab.~\ref{tab:lat_gap}) strongly supports the recent experimental findings 
\cite{Davydov/etal:2002,Wu/etal:2002_1,Nanishi/etal:2003}.
Similar conclusions were drawn from previous $G_0W_0$ calculations applied to 
SIC-LDA ground states, however, only after
adjusting the $pd$ repulsion and combining calculations with and without 
the 4$d$ electrons in the core of the In pseudopotential 
\cite{Bechstedt/Furthmueller:2002,Furthmueller/etal:2005}.

\begin{table}
  \begin{ruledtabular}
    \begin{tabular}{l|ddd|dd}
     param.   & \multicolumn{1}{c}{$a_0$} & 
          \multicolumn{1}{c}{$c_0$} & 
          \multicolumn{1}{c|}{$u$} &
          \multicolumn{1}{c}{$E_{g}$} &
          \multicolumn{1}{c}{$\Delta_1$}\\
     unit & \multicolumn{1}{c}{ \AA} & 
	  \multicolumn{1}{c}{ \AA} & 
          \multicolumn{1}{c|}{} & 
          \multicolumn{1}{c}{eV} & 
          \multicolumn{1}{c}{eV} \\
      \hline
      zb-GaN & 4.50  &       &       & 3.07 &        \\
      zb-InN & 4.98  &       &       & 0.53 &        \\ \hline
      wz-GaN & 3.181 & 5.166 & 0.377 & 3.32 & 0.029  \\
      wz-InN & 3.533 & 5.693 & 0.379 & 0.72 & 0.067  \\
      \hline
    \end{tabular} 
  \end{ruledtabular}
  \caption{\label{tab:lat_gap}\small Lattice parameters, $G_0W_0$ band gap $E_g$,
           crystal field splitting $\Delta_1$ 
	   for zinc blende (zb) and wurtzite (wz) GaN and InN.}
\end{table}

The OEPx calculations in the present work were performed with the plane-wave,
pseudopotential code \texttt{sfhingx} \cite{SFHIngX}, while for the $G_0W_0$ calculations 
we have employed the $G_0W_0$ space-time method
\cite{Rojas/Godby/Needs:1995} in the \texttt{gwst} implementation 
\cite{GW_space-time_method:1998,GW_space-time_method_enh:2000}. 
Exact-exchange pseudopotentials \cite{Moukara/etal:2000} were used throughout 
and the cation $d$-electrons
were  included explicitly \cite{Qteish/etal:2005,Rinke/etal:2005}. 
The calculations
were performed at the experimental lattice constants \cite{upar} taken from
Ref.~\cite{Stampfl/VandeWalle:1999} and
reported in Tab.~\ref{tab:lat_gap}.  
Convergence to within 0.05 eV in the quasiparticle energies 
was achieved for a 4$\times$4$\times$4 (4$\times$4$\times$2)
$\bf k$-point sampling in the zinc blende (wurtzite) phase and a
plane-wave cutoff of 75 (65) Ry for GaN (InN).  
Unoccupied bands up to 55 (45) Ry were included in the calculation of the polarisability for
GaN (InN) in the OEPx as well as in the $G_0W_0$ calculations.

The $G_0W_0$ band gap $E_g$ and the crystal field splitting $\Delta_1$  
for zinc blende and wurtzite GaN and InN are shown in Tab.~\ref{tab:lat_gap}.
Since $G_0W_0$ falls into the realm of perturbation theory it is not {\it a
priori} clear if the quasiparticle corrections are positive or negative. 
While LDA based $G_0W_0$ calculations generally open the band gap from the
underestimated LDA value we observe here that the OEPx gap of 1.0~eV for
wz-InN closes to 0.72~eV after applying the $G_0W_0$ corrections. 
This is not unexpected since the dielectric screening inherent to $G_0W_0$
but not to OEPx counteracts the exchange effects.

Figure \ref{fig:InN-CB} shows the $G_0W_0$ bandstructure for wurtzite InN 
and the conduction band of wurtzite InN and GaN (circles).
To make contact with experimental results we use 
an analytic expression for the conduction band around the $\Gamma$-point
\begin{equation}
\label{eq:CB_kp}
E_c(k)=\frac{\hbar^2 k^2}{2m_0}+\frac{1}{2}
              \left(E_g^{+}+\sqrt{(E_g^{-})^2 
	            +4E_p \frac{\hbar^2 k^2}{2m_0}}\right),
\end{equation}
derived from a four band $\kpe$ model, neglecting spin-orbit splitting.
Here $m_0$ is the free electron mass and
$E_g^{\pm}=E_g\pm\Delta_1$.
The parameter $E_p$ is related to the optical matrix elements between 
conduction
and valence bands. Since it is the only unknown in Eq.~\ref{eq:CB_kp} 
it has been determined by fitting to the $G_0W_0$ conduction band. 
The small reciprocal lattice vector spacing required for an accurate fit 
is easily realised in the
$G_0W_0$ space-time method \cite{Rojas/Godby/Needs:1995}
by means of Fourier interpolation \cite{GW_space-time_method:1998}.

\begin{figure}
  \begin{center} 
    \epsfig{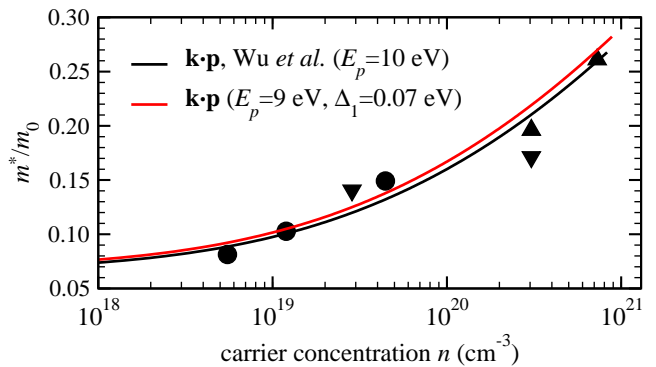}
    \caption{\label{fig:InN-meff} 
             For wz-InN the effective mass as a function of carrier 
	     concentration, deduced
	     from the $G_0W_0$ calculations by means of Eq.~\ref{eq:CB_kp} 
	     and \ref{eq:meff} (red line), agrees well with 
     	     experimental measurements (symbols)
	     and the $\kpe$ fit of Wu {\it et al.} (black line) based on the
	     experimental data
	     \cite{Wu/etal:2002,Walukiewicz/etal:2004}.
	     (Figure adapted from \cite{Walukiewicz/etal:2004})
	     }
  \end{center}
\end{figure}

Using the values of Table \ref{tab:lat_gap} for $\Delta_1$ and $E_g$ we find that the 
quasiparticle conduction band of InN is well described by Eq.~\ref{eq:CB_kp} 
and $E_p^{\parallel}$=$E_p^{\bot}$=9.0 eV
(red solid line in Fig.~\ref{fig:InN-CB}). For GaN
the transition matrix elements differ for different directions and we 
obtain $E_p^{\parallel}$=14.9~eV along the $c$-axis 
and $E_p^{\bot}$=13.6~eV in the basal plane (black solid line).

\begin{figure}
  \begin{center} 
    \epsfig{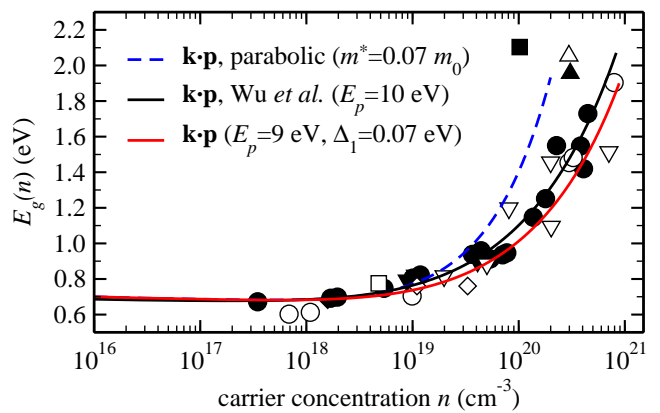}
    \caption{\label{fig:InN-MB} 
             The Burnstein-Moss effect deduced from the $G_0W_0$ band structure (red line)
	     for wz-InN reproduces the experimental trend (symbols) 
             very well and is also close to
	     the $\kpe$ curve of Wu {\it et al} (black line) 
	     \cite{Wu/etal:2002,Walukiewicz/etal:2004}. 
	     Assuming parabolic bands (dashed line) overestimates the
	     Burnstein-Moss shift. (Figure adapted from Ref. 
	     \cite{Butcher/Tansley:2005}; the following symbols differ: 
	     $\bullet$            \cite{Walukiewicz/etal:2004}
	     $\lozenge$           \cite{Losurdo/etal:2006}
	     $\triangle$          \cite{Tyagai/etal:1977})
	     }
  \end{center}
\end{figure}

\begin{table*}
  \begin{ruledtabular}
    \begin{tabular}{l|dd|ddddddd|dd}
     parameter   & 
          \multicolumn{1}{c}{$m_{\bot}^{*}$} &
          \multicolumn{1}{c|}{$m_{\parallel}^{*}$} &
          \multicolumn{1}{c}{$A_1$/$\gamma_1$} &
          \multicolumn{1}{c}{$A_2$/$\gamma_2$} &
          \multicolumn{1}{c}{$A_3$/$\gamma_3$} &
          \multicolumn{1}{c}{$A_4$} &
          \multicolumn{1}{c}{$A_5$} &
          \multicolumn{1}{c}{$A_6$} &
          \multicolumn{1}{c|}{$A_7$} &
          \multicolumn{1}{c}{$E_{p}^{\bot}$/$E_p$} &
          \multicolumn{1}{c}{$E_{p}^{\parallel}$}  \\
     unit &  
          \multicolumn{1}{c}{($m_0$)} & 
          \multicolumn{1}{c|}{($m_0$)} & 
          \multicolumn{1}{c}{} & 
          \multicolumn{1}{c}{} & 
          \multicolumn{1}{c}{} & 
          \multicolumn{1}{c}{} & 
          \multicolumn{1}{c}{} & 
          \multicolumn{1}{c}{} & 
          \multicolumn{1}{c|}{(eV\AA)} & 
          \multicolumn{1}{c}{(eV)} & 
          \multicolumn{1}{c}{(eV)}  \\
      \hline
      zb-GaN & 0.193 &       &   2.506 &  0.636 &  0.977 & & & & & 16.86 &\\
      zb-InN & 0.054 &       &   6.817 &  2.810 &  3.121 & & & & & 11.37 & \\ \hline
      wz-GaN & 0.212 & 0.190 &  -5.798 & -0.545 &  5.259
               & -2.473 & -2.491 & -3.143 & 0.049 & 16.22 & 17.39 \\ 
      wz-InN & 0.071 & 0.067 & -15.230 & -0.520 & 14.673
               & -7.012 & -6.948 & -9.794 & 0.174 & 8.89 &  8.97
    \end{tabular} 
  \end{ruledtabular}
  \caption{\label{tab:par}\small Conduction band effective masses $m^*$,
  Luttinger ($\gamma_1$-$\gamma_3$) and
  $A$ parameters  as well as values 
  for $E_p$ (zinc blende) and
  $E_{p}^{\bot}$ and $E_{p}^{\parallel}$ (wurtzite) \cite{kppar}
  for GaN and InN obtained by fitting a $4\times4$ $\kpe$ Hamiltonian to the $G_0W_0$
  band structure.}
\end{table*}

While the conduction band is well described in 
the effective mass approximation for GaN 
(blue dashed lines in Fig. \ref{fig:InN-CB}) 
it exhibits a deviation from the parabolic shape in InN.
Wu {\it et al.} have recently reported a dependence of the conduction band 
effective mass on the free carrier 
concentration in InN \cite{Wu/etal:2002,Walukiewicz/etal:2004}, which
indicates a pronounced non-parabolicity of the conduction band. 
In the spherical band approximation the momentum effective mass
\begin{equation}
\label{eq:meff}
\frac{m^*(k_F)}{m_{0}}= \left(  \frac{m_0}{\hbar^2 k_F}
                                \left.\frac{dE_c(k)}{dk}\right|_{k=k_F}
	                \right)^{-1}
\end{equation}
can be translated into a carrier concentration dependent effective mass using 
the free electron
relation $k_F=(3\pi^2n)^{1/3}$, where $n$ is the density of electrons in the
conduction band \cite{Wu/etal:2002,Walukiewicz/etal:2004}.
Figure \ref{fig:InN-meff} shows the effective mass of wurtzite InN as a function 
of the free carrier concentration. 
The {\it ab initio} prediction (red line)
extracted from our $G_0W_0$ band structure by means of Eq.~\ref{eq:CB_kp} 
and \ref{eq:meff} reproduces the experimental results very well and closely 
matches the curve obtained by Wu {\it et al.} with an experimentally 
deduced value of $E_p$=10 eV  \cite{Wu/etal:2002,Walukiewicz/etal:2004} 
(black line).

Equation \ref{eq:CB_kp} can also be used to calculate the shift 
of the direct optical transitions ($E_g(n)=E_c(n)-E_v(n)$) towards higher
energies in absorption measurements  
upon increasing electron concentration in the conduction band -- the so
called Burnstein-Moss effect. 
Contributions from the electron-ion and electron-electron 
repulsion at high electron concentrations are accounted for following 
Wu {\it et al.}  \cite{Wu/etal:2002,Walukiewicz/etal:2004}.
The Burnstein-Moss shift calculated in this way fits a wide range of experimentally
reported measurements very well, as shown in Fig. \ref{fig:InN-MB}, and agrees well
with the curve deduced by Wu {\it et al.} from their experimentally determined
values of $E_g$ and $E_p$. Neglecting the non-parabolicity of the conduction 
band (blue dashed line)
worsens the agreement with the experimental results.

For device simulations \mbox{$\kpe$} models have been well established. 
However, so far most of the parameters entering the $\kpe$ Hamiltonian 
\cite{Kane,Chuang/Chang:1996} for GaN and InN have not been conclusively 
determined by experiment \cite{Vurgaftman/Meyer:2003}. We have therefore used 
our quasiparticle band structure to derive a complete set of 
 Luttinger and $A$-parameters for zinc blende and wurtzite InN and GaN \cite{fit}.   
The parameters \cite{kppar} are shown in Tab.~\ref{tab:par}. 

In conclusion, we have carried out $G_0W_0$ band structure calculations based on
exact-exchange density functional theory for InN and GaN. From these results 
we have derived key electronic quantities (band gaps, carrier dependence of 
the effective mass, $\kpe$ parameters). For InN, a band gap of $\approx 0.7$\,eV 
is found supporting recent experimental observations. 
For all investigated parameters an excellent agreement with experimental data has 
been observed, indicating that the wide interval of experimentally observed 
band gaps can be largely explained by the Burnstein-Moss effect.

We would like to acknowledge fruitful discussions with Peter Kratzer,
Matthias Wahn and Christoph Freysoldt.
This work was in part supported by the Volkswagen Stiftung/Germany, the DFG
through Sfb 296 and the research group 
,,nitride based nanostructures" and
the EU's 6th Framework Programme through the 
NANOQUANTA (NMP4-CT-2004-500198) and 
SANDiE (NMP4-CT-2004-500101) Networks of Excellence.

\bibliographystyle{apsrev}

\begin{thebibliography}{32}
\bibitem[{upa()}]{upar}
\bibinfo{note}{For the wurtzite phases the internal lattice parameter $u$ was
  determined by minimising the total energy in the LDA at fixed lattice
  constants $a_0$ and $c_0$.}
\bibitem[{kpp()}]{kppar}
\bibinfo{note}{Note that Eq.~\ref{eq:CB_kp} results from a simplification of
  the $\kpe$ Hamiltonian used for fitting the $A$-parameters. Therefore the
  values for $E_{p}$ differ from Fig.~\ref{fig:InN-CB} to Tab.~\ref{tab:par}.}
\bibitem[{fit()}]{fit}
\bibinfo{note}{The fit is not restricted to parabolic bands and $E_{p}^{\bot}$
  and $E_{p}^{\parallel}$ are determined independently. Details of the
  procedure will be published elsewhere.}
\end{thebibliography}

\end{document}